%% file: TC_paper.tex
\documentclass[10pt, conference]{IEEEtran}
\IEEEoverridecommandlockouts
\usepackage{cite}
\usepackage{amsmath,amssymb,amsfonts}
\usepackage{graphicx}
\usepackage{textcomp}
\usepackage[dvipsnames]{xcolor}
\usepackage{url}
\usepackage{booktabs,rotating}
\usepackage{multirow}
\usepackage[noend]{algpseudocode}
\usepackage{algorithm}

\usepackage{subcaption}
\usepackage{xcolor, colortbl, hhline}
\usepackage{nicematrix}
\usepackage{layouts}

\usepackage{footnote}
\usepackage{enumitem}
\makesavenoteenv{tabular}
\makesavenoteenv{table}

\definecolor{pastelblue}{rgb}{0.68, 0.78, 0.81}
\definecolor{pastelgreen}{rgb}{0.47, 0.87, 0.47}
\definecolor{green}{RGB}{0,138,0}

\newtheorem{observation}{Observation}[section]

\newcommand{\salvo}[1]{\textcolor{red}{}}
\newcommand{\andris}[1]{\textcolor{blue}{}}
\newcommand{\fv}[1]{\textcolor{green}{}}

\usepackage{tikz}
\usetikzlibrary{matrix}

\newcommand*\negcircnum[1]{\tikz[baseline=(char.base)]{%
            \node[white,shape=circle,fill=green,draw,inner sep=1pt] (char) {\color{white}\sffamily\small\textbf{#1}};}}

\def\BibTeX{{\rm B\kern-.05em{\sc i\kern-.025em b}\kern-.08em
    T\kern-.1667em\lower.7ex\hbox{E}\kern-.125emX}}

\usepackage{marginnote}
\usepackage{soul} 
\usepackage{xcolor}

\DeclareDocumentCommand\review{m g g} {#1}

\begin{document}

\title{Asynchronous Distributed-Memory Triangle Counting and LCC with RMA Caching

}

\author{\IEEEauthorblockN{András Strausz\IEEEauthorrefmark{1},
                        Flavio Vella\IEEEauthorrefmark{2},
                        Salvatore Di Girolamo\IEEEauthorrefmark{3}, 
                        Maciej Besta\IEEEauthorrefmark{3}
                        and Torsten Hoefler\IEEEauthorrefmark{3}}
        \IEEEauthorblockA{\IEEEauthorrefmark{1}\IEEEauthorrefmark{3}Dept. of Computer Science, ETH Zürich, Zürich, Switzerland}
        \IEEEauthorblockA{\IEEEauthorrefmark{2}Dept. of Engineering and Computer Science, University of Trento, Trento, Italy\\
        \IEEEauthorrefmark{1}strausza@student.ethz.ch, \IEEEauthorrefmark{2}flavio.vella@unitn.it, \IEEEauthorrefmark{3}firstname.lastname@inf.ethz.ch}
}

\maketitle

\begin{abstract}
Triangle count and local clustering coefficient are two core metrics for graph analysis. They find broad application in analyses such as community detection and link recommendation.
%
To cope with the computational and memory demands that stem from the size of today's graph datasets, distributed-memory algorithms have to be developed. 
%
Current state-of-the-art solutions suffer from synchronization overheads or expensive pre-computations needed to distribute the graph, achieving limited scaling capabilities.
%
We propose a fully asynchronous implementation for triangle counting and local clustering coefficient based on 1D partitioning, using remote memory accesses for transferring data and avoid synchronization. 
Additionally, we show how these algorithms present data reuse on remote memory accesses and how the overall communication time can be improved by caching these accesses. Finally, we extend CLaMPI, a software-layer caching system for MPI RMA, to include application-specific scores for cached entries and influence the eviction procedure to improve caching efficiency.
%
Our results show improvements on shared memory, and we achieve 14x speedup from 4 to 64 nodes for the LiveJournal1 graph on distributed memory. Moreover, we demonstrate how caching remote accesses reduces total running time by up to \reversemarginpar \review{73\%}{G1}{Green} \normalmarginpar with respect to a non-cached version. Finally, we compare our implementation to TriC, the 2020 graph champion paper, and achieve up to \review{100x}{G2}{Green} faster results for scale-free graphs.

\end{abstract}

\vspace{-1em}
\section{Introduction}

Complex real-world systems can be modeled, analyzed, and optimized through their respective graph representations. 
These systems range from social networks~\cite{tang2010graph}, to biological networks~\cite{aittokallio2006graph}, to the full Internet~\cite{boccaletti2006complex}.
Data mining, information retrieval, recommendation systems, and fraud detection are just a few applications of graph analysis~\cite{cook2006mining, besta2021graphminesuite, gianinazzi2021parallel, benson2016higher, besta2021sisa}.

The local clustering coefficient (LCC)~\cite{watts1998collective} indicates the likelihood that neighbors of a node are also neighbors to each other, which has applications in the link prediction problem~\cite{lu2011link}. 
In particular, this metric has been proven useful in many applications, such as community detection~\cite{communitydetection, besta2020communication} or link recommendation~\cite{linkrecommendation}.
%
%
In the former case, LCC is used to detect communities in, e.g., social networks, distinguishing between vertices that are central to the cluster from others on its frontier. 
%
In the latter, clustering coefficient is used to locate thematic relationships by looking at the graph of hyperlinks.

\begin{figure}[t]
\centering
\includegraphics[width=0.9\columnwidth]{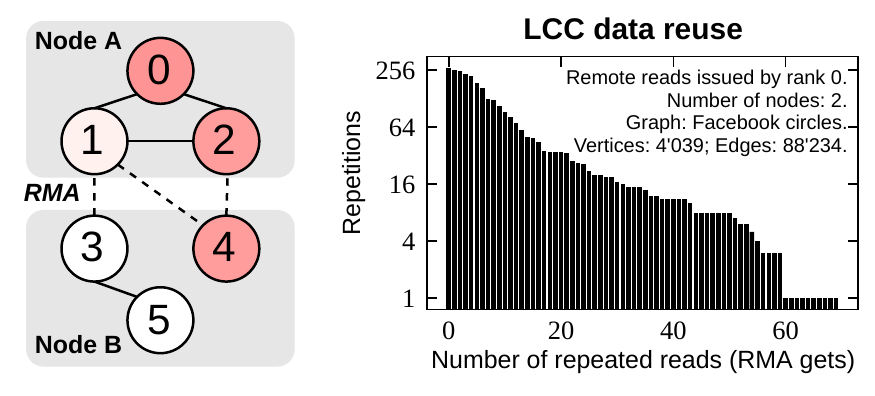}
\vspace{-0.5em}
\caption{LCC example and data reuse. Left: graph partitioned between two compute nodes. The gradient of the vertices indicates their LCC score. Dashed edges indicate RMA communications needed to read adjacency lists from remote nodes. Right: data reuse in social network graph~\cite{mcauley2012learning}.}
\vspace{-1.5em}
\label{fig:intro:overview}
\end{figure}

The LCC of a vertex is given as the fraction of the pairs of its neighbors that are themselves connected by an edge. 
Figure~\ref{fig:intro:overview} (left) shows an example of LCC scores on a toy graph. Vertices with darker backgrounds have higher LCC.
Given a vertex, two of its neighbors contribute to its LCC score if the three vertices form a triangle in the graph. Therefore, to compute the LCC, one has to count the number of triangles that are closed by its neighbors for every vertex.

With the rapid growth in the size of graphs to be analyzed, both memory and computational capacities of a single machine become insufficient to perform triangle counting (TC) analysis on a single node. There are two main strategies to compute TC in distributed memory: (1) by first computing overlapping partitions that are necessary for local triangle counting or (2) by issuing communications between processes. Current state-of-the-art solutions~\cite{TRUST,hu2018tricore,Tric,hoang2019disttc} all follow either the Bulk Synchronous Parallel model~\cite{BSP} or MapReduce~\cite{kolda2014counting} and suffer from synchronization, as well as the overhead of computing the partitioning of the graph. The 2020 graph champion paper TriC~\cite{Tric} utilizes blocking all-to-all communication resulting in synchronization overheads being as costly as communication. To minimize communication, DistTC~\cite{hoang2019disttc} computes and distributes shadow edges that are necessary for computing triangles locally. This approach leads to a low computation time but makes the total running time dominated by this pre-computation step, similarly limiting scalability.

However, distributed triangle counting and local clustering coefficient do not necessarily require synchronization. In fact, as the graph is not updated during the computation, the distributed algorithm can be organized to let different processes progress asynchronously while still storing only partitions of the graph. We exploit this characteristic by proposing a fully asynchronous solution that uses Remote Memory Access (RMA) one-sided operations to read remote portions of the graph without involving target nodes. 
For example, the graph of Figure~\ref{fig:intro:overview} (left) is distributed on two computing nodes. To compute the LCC of vertex 1, node A reads the adjacency lists of vertices 0 and 2 locally and the adjacency list of vertex 4 via RMA. 

We notice how some RMA accesses are repeated and, in principle, can be avoided to save communication time. For example, when node A needs to compute the LCC of vertex 2, it will again read via RMA the adjacency list of vertex 4. 
Figure~\ref{fig:intro:overview} (right) shows the data reuse in a real-world graph modeling social network circles~\cite{mcauley2012learning}. The plot shows how many remote reads ($x$-axis) are repeated $y$ times when the graph is partitioned among two computing nodes. 
We exploit data reuse in the remote access pattern of LCC computations by caching RMA accesses using CLaMPI~\cite{rma-caching}, a transparent caching layer for RMA. Moreover, we extend CLaMPI to accept application-defined scores for cached entries and show how this can improve caching efficiency.

All in all, in this work we:
\begin{itemize}[noitemsep,topsep=1pt,parsep=0pt,partopsep=0pt,leftmargin=10pt]
	\item propose a distributed and fully asynchronous algorithm for both triangle counting and LCC (Section~\ref{sec:distlcc}).
    \item introduce a hybrid strategy for triangle computation based on the frontiers (Section~\ref{sec:dynamic}).
    \item show how data reuse in remote access patterns of LCC computation can be exploited by caching RMA accesses and reduce the overall communication time. We further increase caching performance by introducing application-defined scores for victim selection (Section~\ref{sec:datareuse}).
\end{itemize}
Our hybrid approach for the local computation of triangles can improve performance by up to 8\% on a CPU. We achieve shared memory parallelism by computing the intersection in parallel using OpenMP, leading to a speedup of up to 2.7$\times$ using 16 threads compared to a sequential implementation. On distributed memory, our non-cached algorithm achieves a speedup of up to  $14\times$. Moreover, we can reduce the total running time using caching by up to \review{$73\%$}{G1}{Green} compared to a non-cached version until the graph is not over-partitioned. Finally, we show up to \review{$100\times$}{G2}{Green} better performance compared to TriC.

\section{Background}
\subsection{Notation}
We denote an unweighted graph that contains no multi-edges and loops as $G = (V, E)$, where $V$ is the set of vertices and  $E \subseteq V \times V$ the set of edges ($|V|=n$ and $|E|=m$). We will use $v_i$ to denote a vertex and $e_{ij}$ for the edge from $v_i$ to $v_j$. We call the adjacency of $v_i$ the set of vertices $adj(v_i) = \{v_j : e_{ij} \in E\}$ and denote by $\mathbf{A}$ the adjacency matrix of $G$. Moreover, we define the in-degree of $v_i$ as $deg^-(v_i) = \left|\{v_j : e_{ji} \in E\}\right|$ and the out-degree as $deg^+(v_i) = \left|\{v_j : e_{ij} \in E\}\right|$ (note, for a directed graph in and out-degree equals). We use the symbol $\triangle_{ijk}$ for the triangle that consists of the edges $e_{ij}$, $e_{ik}$ and $e_{jk}$. We will denote by $p$ the number of processes. For simplicity, we assume that $p$ is always a power of 2 and $p$ divides $n$. A summary of the notation used in the paper can be found in Table~\ref{tab:symbols}.

\subsection{Graph format}\label{sec:graph_format}
\input{Inlines/symbols}
We consider graphs with no multi-edges and remove vertices that have degree less than two, as they cannot be part of any triangle. If the input graph is stored in a degree-ordered format, we use a random relabeling to avoid assigning all the highest degree vertices to the same process.

\begin{figure}[H]
    \vspace{-1em}
    \centering
    \includegraphics[width=0.75\columnwidth]{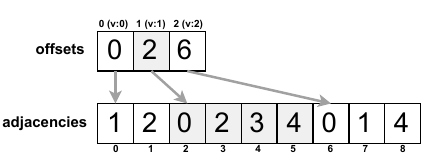}
    \caption{\review{Compressed Sparse Row (CSR) representation of the subgraph stored on node A of Figure~\ref{fig:intro:overview}.}{R4.1}{red}}
    \label{fig:csr_example}
\end{figure}

The graph is stored in the CSR (Compressed Sparse Row) format (see Figure~\ref{fig:csr_example}), where each process stores its partition with the help of two arrays: \texttt{offsets} and \texttt{adjacencies}. 
An element $i$ of the \texttt{offsets} array stores the offset at which adjacency list of vertex $v_i$ starts in the \texttt{adjacencies} array.

\subsection{Triangle computation}
\label{sec:tc}

We follow the edge-centric method for triangle counting and compute the number of triangles that are closed by an edge $e_{ij}$ for every $e_{ij}\in E$. This is given by the number of common neighbours, which can be formulated as $\left| adj(v_i) \cap adj(v_j) \right|$. We note that in the undirected case, the presence of a triangle  $\triangle_{ijk}$ implies the presence of $\triangle_{ikj}$. However, in the general edge-centric method, both triangles are enumerated because they lie on different edges. To reduce computation, this double counting can be eliminated by offsetting the neighbor's adjacency to count the common elements only for the upper triangle in the adjacency matrix, that is, for the set of vertices $\{v_k : v_k\in adj(v_j) \wedge j < k$\}.

For the computation of the intersection, we apply either binary search or sorted set intersection (SSI), which proceed for two sorted lists $A$ and $B$ with $|A| \leq |B|$ as follows:
 
\subsubsection{Binary search}
With binary search, the computation of the intersection breaks down to issuing $|A|$ lookups in a sorted array of length $|B|$ resulting in a \review{running time complexity}{R4.2}{red} of $\mathcal{O}(|A| \cdot log(|B|))$. To minimize this, one should always assign the longer list as the search tree and the shorter one as the array of keys.
  
\subsubsection{Sorted Set Intersection}
SSI traverses the two lists simultaneously by comparing the current elements and progressing the array whose current element is smaller. If a common element is found, it increments the counter and moves an element in both arrays. Trivially, SSI computes the intersection of two lists in $\mathcal{O}(|A| + |B|)$.
 
The described algorithms are outlined in Algorithm~\ref{algo:bin_search} and \ref{algo:ssi}. We emphasize that these methods assume sorted adjacency lists, however, most graph datasets are already of this form.

\input{Inlines/bin_search}
\input{Inlines/ssi}

\subsection{Local Clustering Coefficient}
The Local Clustering Coefficient of a vertex $v_i$ was defined by Watts and Strogatz~\cite{watts1998collective} as the proportion of existing edges between the vertices adjacent to $v_i$ divided by the possible number of edges that can exist between them. For directed graphs, this can be computed as:
\begin{equation}
\label{eq:lcc-directed}
 C(i) = \frac{|\{e_{jk}: v_j, v_k \in adj(v_i), e_{jk} \in E\}|}{deg^+(v_i) \cdot (deg^+(v_i) - 1)}
\end{equation}
Similarly, for an undirected graph:
\begin{equation}
\label{eq:lcc-undirected}
 C(i) = \frac{2 \cdot |\{e_{jk}: v_j, v_k \in adj(v_i), e_{jk} \in E\}|}{deg^+(v_i) \cdot (deg^+(v_i) - 1)}
\end{equation}
For a pair of vertices $\{v_j, v_k\}$, in order to contribute to the numerator, the edges $e_{ij}$, $e_{ik}$ and $e_{jk}$ must exist, forming the triangle $\triangle_{ijk}$ in $G$. Thus, if vertex degrees are known, LCC can be computed by detecting, for every vertex, the number of triangles in which they participate.

\subsection{MPI-RMA}

Remote Memory Access (RMA) operations are defined by the MPI-3 standard~\cite{hoefler2015remote} and enable MPI processes to access memory regions of remote peers in a one-sided fashion. When running on networks supporting remote direct memory access (RDMA), RMA operations are naturally mapped to the hardware interface (e.g., ibverbs~\cite{bedeir2010rdma}, uGNI~\cite{alverson2012cray,uGNI2}), resulting in higher throughput and lower latency. 
\review{While also two-sided communications can benefit from a RDMA-based implementation, they still incur in overheads caused by MPI message matching that can lead to additional message copies or synchronization. }\review{For this reason, in this work we focus on MPI RMA.}{R3.7}{orange}
Processes can expose their local memory by creating a window that serves as a logically distributed memory region. To access remote memories, MPI processes can use functions like \texttt{MPI\_Put} and \texttt{MPI\_Get}. With the first, a process can write into memory regions exposed over the network by remote peers. With the second, a process can read the content of such regions. Communications in MPI RMA are always non-blocking, and synchronization is enforced only at the beginning and at the end of an epoch. The process-local memory region can be accessed by other processes during an exposure epoch, whereas a process can access remote data during an access epoch. MPI defines two types of synchronization modes: active and passive. With the first, both initiator (i.e., the process issuing RMA operations) and target (i.e., the process targeted by RMA operations) processes synchronize to start a new epoch. With the passive mode, the participation of the target process is not required.

\begin{figure*}[t]
\includegraphics[width=\textwidth]{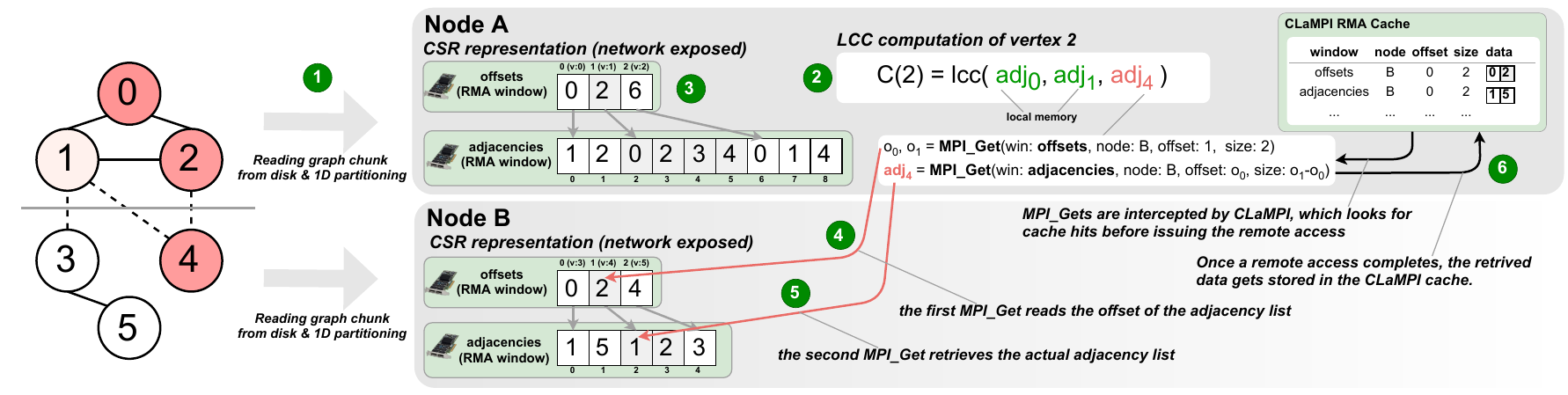}
\caption{\review{Overview of the proposed approach for distributed, fully-asynchronous LCC computation with RMA caching.}{R4.1}{red}}
\label{fig:full-overview}
\end{figure*}

\subsection{RMA Caching}
CLaMPI~\cite{rma-caching} is a software caching layer that transparently caches data retrieved through MPI RMA operations. CLaMPI can be linked to MPI applications, and it is designed to fit into the \mbox{MPI-3 RMA} programming model, minimizing cache-management overheads and relying on the concept of MPI epochs to enforce consistency.

As applications can issue arbitrary-size read operations, CLaMPI supports caching of variable-size entries. This is achieved by using two data structures: a hash table to index cached entries and an AVL tree to store free regions in the memory buffer reserved for caching. Both the size of the hash table and the capacity of the memory buffer are parameters that can be tuned to the specific use case. 
CLaMPI includes an adaptive parameter tuning heuristic that automatically resizes the hash table and the memory buffer by observing indicators such as cache misses, conflicts in the hash table, and evictions due to lack of space in the memory buffer.

In CLaMPI, the eviction procedure is triggered when the application makes remote memory accesses over an MPI window that: (a) is enabled for caching, (b) does not contain the referenced data in cache, and (c) does not have enough space either in the hash table or in the memory buffer to index or store the new data. Due to the variable size of the cached entries, the system can incur in external fragmentation of the memory buffer: the free space can be fragmented in small non-contiguous regions that cannot fit a new entry. To reduce external fragmentation, CLaMPI assigns a score to the cached entries that reflect both their temporal locality and how much fragmentation they are causing in the memory buffer. The victim selection takes this score into account when deciding which entry to evict: e.g., if an entry is poorly placed in the memory buffer (e.g., surrounded by free space that could be merged if the entry would be removed), it will be more likely to be evicted even if it presents higher temporal locality. 

CLaMPI provides three operational modes to enforce consistency of cached data: \emph{transparent}, \emph{always-cache}, and \emph{user-defined}. The \emph{transparent} mode does not make any assumption on the nature of the cached data (e.g., read-only or read-write) and flushes the cache at every epoch closure. In this case, CLaMPI can still save remote accesses that target the same data and that are made within the same epoch. However, cached data does not persist among multiple epochs. Consistency within the epoch is guaranteed by the MPI RMA semantic that, e.g., forbids conflicting \emph{put} and \emph{get} operations happening in the same epoch. The \emph{always-cache} mode assumes that data accesses with RMA operations is read-only. In this case, there is no need to flush the cache since there are no updates to be propagated. Finally, the \emph{user-defined} mode leaves the responsibility of flushing the cache to the application. For example, an application might be using data as read-only for a number of epochs, during which cached data can persist, and then switch to another phase where updates are issued, hence needing to flush the cache.

\section{Accelerating Distributed LCC}

LCC can be formulated in terms of counting the number of closed triplets centered in a node. This formulation enables the use of triangle counting as a fundamental primitive for LCC computation. Moreover, it implies that it is possible to compute LCC of each vertex asynchronously. The problem of computing triangles in large-scale graphs has been widely investigated, and, as we mentioned, the main limitation to scalability comes from the synchronization cost and the unbalancing of the graph partitioning. 


In our algorithm design, we explore asynchronous computation and a mechanism for improving data locality based on the concept of vertex delegation.

To this aim, we distribute the input graph among multiple computing nodes and let them access remote partitions via one-sided RMA operations. By distributing the graph, we lower the per-node memory requirements and reduce initialization overheads (i.e., I/O time to read the graph), which would otherwise limit strong-scaling capabilities because of Amdahl's law. Additionally, we show how the remote memory access pattern of LCC presents data reuse (temporal locality), which we exploit by caching RMA \emph{get} operations with CLaMPI. Figure~\ref{fig:full-overview} shows an overview about the proposed approach.
\input{Inlines/lcc_comp}

\subsection{Asynchronous computation}
\label{sec:distlcc}

We use a 1D partitioning scheme to distribute the graph among the processes~\negcircnum{1}. In this scheme, an equal number of vertices are assigned to each process. 
With $p$ computing nodes this is given by $V = V_1 \cup V_2 \ldots \cup V_p$, such that:
\begin{displaymath}
V_k = \left\{v_i : i \in \left(\frac{(k-1) \cdot n}{p}, \frac{k \cdot n}{p}\right]\right\}
\end{displaymath}
We note that, in case the degrees are highly skewed, this partitioning method can introduce load imbalance between processes. A more balanced partitioning can be achieved by cyclic distribution~\cite{CyclicDistribution}. 
However, this would require sorting vertices, introducing additional computation as well as communication.

Processes compute the number of triangles for every locally owned vertex~\negcircnum{2}. In the edge-centric method (see Section~\ref{sec:tc}), a process $p_i$ computes $|adj(v_i) \cap adj(v_j)|$ for every $e_{i,j} \in E$ such that $v_i \in V_i$ and $v_j \in adj(v_i)$. In case the vertices $v_i$ and $v_j$ belong to different partitions, the owner process first reads the adjacency of $v_j$ from remote memory. In the CSR representation, the degree of a vertex is implicitly stored in the \texttt{offsets} array, therefore, after computing the number of triangles the LCC score is instantly attainable. The algorithm is outlined in Algorithm~\ref{algo:dist_lcc}.

As the graph is stored in CSR representation~\negcircnum{3}, to read the adjacency list of a vertex from a remote node, we need to perform two remote read operations: one for reading from the \texttt{offsets} array the offset of the adjacency list in the \texttt{adjacencies} array~\negcircnum{4}, followed by a second one that reads the actual adjacency list from the \texttt{adjacencies} array (i.e., starting at the right offset)~\negcircnum{5}. 

To enable remote access, processes expose their local graph partitions over the network. 
As a result, the graph is logically shared among all the processes: it can be accessed either locally (i.e., for locally owned partitions) or remotely (i.e., for partitions owned by other processes).
Using MPI-RMA, this can be achieved in two windows, denoted by $w_{\textrm{offsets}}$ and $w_{\textrm{adj}}$~\negcircnum{3}, in which processes share their local \texttt{offsets} and \texttt{adjacencies} arrays, respectively. When performing remote reads, processes first issue a RMA \emph{get} targeting the $w_{\textrm{offsets}}$ RMA, then issue a RMA \emph{get} on the $w_{\textrm{adj}}$ window to retrieve the adjacency list.
To guarantee that the algorithm is fully asynchronous, we adopt the MPI passive target synchronization mode~\cite{hoefler2015remote}. In this way, processes initially expose the interested memory regions in RMA windows, which then become accessible from remote peers without further synchronization. A process that wants to access a remote window first calls a \texttt{MPI\_Win\_lock\_all} to start an access epoch. This is followed by one or more RMA \emph{get} operations. \emph{We remark on the unfortunate name of \texttt{MPI\_Win\_lock\_all}: it is not an actual lock, and thus it does not synchronize processes. Instead, its effect is to signal the beginning of a new access epoch; the remote window can still be accessed by multiple (``all'') processes.}
After accessing data, the process can then perform a \texttt{MPI\_Win\_flush} operation or close to access epoch with an \texttt{MPI\_Win\_unlock\_all} to make sure that the remote reads are completed and that the relative data can be accessed without risk of corruption. Even in this case, the closure of an access epoch in the passive synchronization mode is a local operation, and it does not require synchronization.

Finally, to increase efficiency, we use a double-buffering approach where we overlap the processing of two consecutive edges by overlapping the computation phase of the current edge with the communication corresponding to the next one.

\subsection{Exploiting data reuse}
\label{sec:datareuse}

While remote memory accesses enable a global memory abstraction where each node can access the memory of remote ones, these accesses are normally one or more orders of magnitude more expensive than accessing local memory. For example, they can take up to 2-3 microseconds on a Cray Aries network~\cite{rma-caching}. In contrast, a DRAM accesses takes hundreds of nanoseconds that become tens of nanoseconds if the data is in cache.

By distributing the graph among multiple processes, a large portion of the edges has endpoints in distinct partitions, thus requiring remote communications. For example, in an R-MAT graph with $2^{20}$ vertices and $2^{24}$ edges equally partitioned among 8 processes using 1D partitioning, 95\% of the edges go between different partitions.

\begin{figure}[t]
\centering
\input{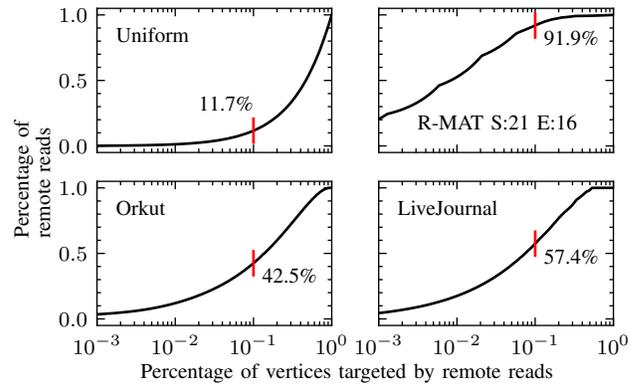}
\caption{Data reuse in four datasets using 8 processes and 1D partitioning. The plot shows how the highest degree vertices contribute to the total number of remote reads issued. Upper left, we show a graph with uniform degree distribution, while the other graphs follow a power law-like degree distribution (see Table~\ref{tab:snap_graphs}). We highlight the fraction of remote reads that target the top 10\% of the highest degree vertices.}
\vspace{-1.5em}
\label{fig:data_reuse}
\end{figure}

Assuming that vertices are randomly assigned to computing nodes, the in-degree of a vertex directly correlates with the number of times it will be remotely accessed. Using $p$ computing nodes, a node will access a non-local vertex $v_j$ in expectation $\frac{deg^-(v_j) - p}{p}$ times. 
Most real-world graphs present a degree distribution that follows a power law. Hence, a large portion of the remote reads will target the same small set of vertices. An illustration of how the highest degree vertices contribute to the number of remote communications can be found in Figure~\ref{fig:data_reuse}.
\salvo{explain fig 3 in the text, make example how to read it}

We exploit temporal locality in the remote memory access pattern by using a caching layer for RMA, CLaMPI, that allows to transparently cache MPI RMA accesses~\negcircnum{6}. 
By caching remote data, each node stores a dynamically defined sub-graph containing vertices that are frequently accessed and thus are expected to be accessed in the future as well. 

As discussed in Section~\ref{sec:distlcc}, an access of a remote adjacency list is done through two steps: a first RMA \emph{get} to read from the $w_{\textrm{offsets}}$ of a remote node, followed by a second RMA \emph{get} to the $w_{\textrm{adj}}$ of the same node (using the data offset read with the first \emph{get}). We enable caching for both RMA windows at every process. This results in two CLaMPI caches: $C_{\textrm{offsets}}$ and $C_{\textrm{adj}}$. \review{The former caches data offsets telling the position at which the adjacency list of a vertex starts and ends}{R3.11}{orange}, \review{whereas the latter stores full adjacency lists of cached vertices.}{R3.4}{orange} Both caches are set to \emph{always-cache} as the graph is never modified during the computation. In this configuration, CLaMPI does not automatically flush caches between access epochs.

\subsubsection{Cached windows characterization}

\begin{figure}[H]
    \vspace{-0.5em}
    \centering
    \includegraphics[width=\columnwidth]{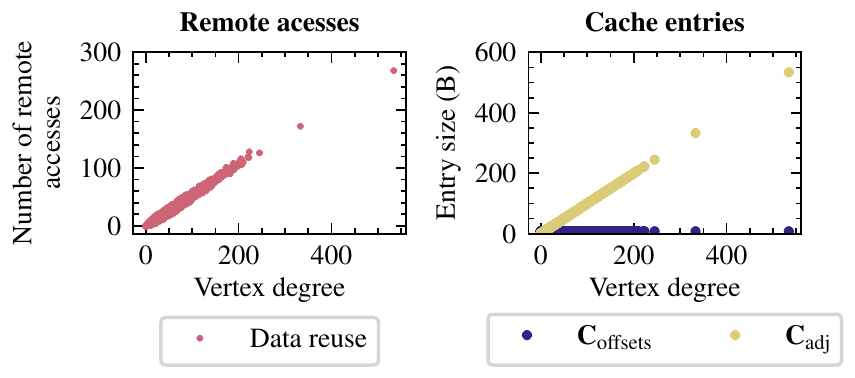}
    \caption{Data reuse and cache entry sizes for the Facebook circles dataset~\cite{mcauley2012learning} distributed among two computing nodes.}
    \label{fig:degree}
    \vspace{-0.5em}
\end{figure}

The following two observations describe the characteristics of the two CLaMPI caches and serve as a ground for their analysis:
\begin{observation}
\label{obs:1}
As noted earlier, the number of accesses to a vertex correlates with its degree. As the entries in $C_{\textrm{adj}}$ are adjacency lists and the size of the adjacency list of a vertex is equal to its degree, the size of the cached adjacency lists is a good indicator for the reuse of entries in $C_{\textrm{adj}}$.
\end{observation}
\begin{observation}
\label{obs:2}
On the other hand, $C_{\textrm{offsets}}$ stores fixed-size entries, namely the offsets of adjacency lists in the remote \texttt{adjacencies} arrays. Therefore, there is no connection between the size of an entry and its reuse. However, an entry that stores the position of the adjacency list of a high-degree vertex will be accessed more often than an entry corresponding to a low-degree vertex.
\end{observation}
We illustrate these observations in Figure~\ref{fig:degree} using the Facebook circles dataset~\cite{mcauley2012learning}. The figure shows how, in the case of $C_{\textrm{adj}}$, the entry reuse (left) correlates with the entry size (right).
Even though for directed graphs in- and out-degree may differ, we expect that the second observation also holds in the directed case due to the high reciprocity in real-world graphs.

We enable CLaMPI's adaptive strategy for the auto-tuning of the hash table size. 
However, as the adaptive strategy flushes the cache every time it adjusts the hash table size, it is crucial to determine good starting values. 
The size of the hash table should be equal to the expected number of entries in the cache. $C_{\textrm{offsets}}$ stores fixed-size entries, where every entry corresponds to one vertex in the graph. Therefore, the number of entries in $C_{\textrm{offsets}}$ is linear in $n$ and in the size of memory allocated for the cache. For example, if the cache size equals $n/2$ bytes, the optimal size of the hash table for $C_{\textrm{offsets}}$ will roughly equal $n/2$.
We expect to cache some of the highest degree vertices. Thus a small number of entries in $C_{\textrm{adj}}$ is likely to take up much of the space allocated for this cache (recall the size of the entry equals the degree of the corresponding vertex).
If the graph's degree distribution follows a power law, the size of the hash table for $C_{\textrm{adj}}$ will be a function of the graph size and the cache size, which will also follow a power law. In this case, if the cache size is half of the graph's size, we expect to store $n \cdot 0.5^{\alpha}$ many entries in the cache. We found that $\alpha = 2$ results in a good approximation for the hash table size.

\subsubsection{Application-defined scores for cached entries}
\review{By default, CLaMPI selects entries to evict with a least recently used (LRU) scheme weighted on a positional score to limit external fragmentation. With this scheme, high-degree vertices can still be evicted if the CLaMPI cache fills up (e.g., due to many low-degree vertices being accessed). 
Additionally, as CLaMPI caches a missing entry only if it has resources to store it, evicted high-degree vertices have a lower chance of being re-cached due to their larger sizes.}

\review{We notice how the vertex degree represents a good indication of the importance of storing that vertex: the higher the degree, the more probable it is that the vertex will be accessed multiple times.}{R3.6}{orange} In particular, Observation~\ref{obs:1} enables us to use this application-specific knowledge about an entry's value to increase further caching performance. By controlling the eviction procedure based on the degree centrality of the vertices, we expect to avoid storing a high number of low-degree vertices. Lower degree vertices would consume space in the cache while having a lower likelihood of being reused.
We modified CLaMPI to accept an application-specific score passed by the user. This score is used by CLaMPI in the victim selection process whenever an entry must be evicted. After completing the \emph{get} targeting $w_s$, we know the out-degree of the non-local vertex, and we can assign it as a score of the respective adjacency list in $C_d$. We note that this extra knowledge about an entry's value may lead to increased performance, but we lose the spatial effect of the score that attempted to reduce fragmentation. 

\subsection{Optimization of local computation}
\label{sec:dynamic}
We take advantage of parallelism on the edge level and compute the intersection of the adjacency lists in parallel with OpenMP. For the binary search-based method, we distribute work among the threads by splitting the shorter (keys) array into equal-sized chunks. On the other hand, for SSI, we split the longer array and every thread intersects the assigned chunk with the shorter list. 
By using parallelism for the computation of the intersection and not on a higher level (e.g. distributing edges among threads) we can achieve low imbalance between the threads. 
However, as a too-small parallel region would limit performance, we determine a cut-off value, for which case the intersection is computed sequentially. Moreover, to decrease the cost of entering a parallel block, we specify \texttt{OMP\_WAIT\_POLICY=active} that forces threads to spin even if they are inactive.

Based on the time complexities of the algorithms described in Section~\ref{sec:tc}, for two sorted lists $A$ and $B$ with $|A| \leq |B|$, one can arrive at the following rule for the case where SSI is theoretically faster:
\begin{equation}
    \frac{|B|}{|A|} \leq log_2(|B|) - 1
    \label{eq:bin_ssi_rule}
\end{equation}
We utilize this decision rule to arrive at a hybrid method for triangle counting where frontiers are empirically compared to decide which method to apply for computing the intersection.

\section{Experimental Evaluation}

\subsection{Experimental setup}

To evaluate our distributed LCC solution, we used R-MAT~\cite{chakrabarti2004r} synthetic graphs and real-world graphs from \review{several databases}{R4.4}{red}~\cite{snapnets, konect, UbiCrawl}. An R-MAT graph with scale $x$ and edge factor $y$ includes $2^x$ vertices and $2^{x+y}$ edges. We generate R-MAT graphs with the parameters $a=0.57$, $b=c=0.19$ and $d=0.05$ for controlling the degree distribution. \reversemarginpar \review{Table~\ref{tab:snap_graphs} lists properties of the graphs that were used for the final experiments and shows the size of their CSR representation after one-degree removal.}{R4.5}{red} \normalmarginpar

Shared memory benchmarks were run on an Intel® Xeon Gold 6154 @ 3.00GHz CPU with 16 cores, and the code was compiled with Intel's ICC 2021.1 with the \texttt{-O3} flag. The distributed version was tested on the Piz Daint cluster at the Swiss National Supercomputing Centre (CSCS). We used the XC50 computing nodes, which are powered by 12 core Intel® Xeon® E5-2690 v3 2.60GHz CPUs equipped with 64GB RAM per node and interconnected with Cray's Aries network arranged in a dragonfly topology. The code was compiled with the ICC 19.1 compiler with \texttt{-O3} flag and using the cray-mpich 7.7.16 MPI implementation.

\review{We distinguish small-scale experiments with no more than 64 computing nodes that we allocate on a single electrical group and large-scale experiments with 128 or more, allocated freely over the whole cluster.}{R3.12}{orange}

Time measurements were taken using the LibLSB library~\cite{benchmarking}. For shared memory experiments, we report the median and repeated every experiment until the 5\% of the median was within the 95\% CI. For distributed memory experiments, we measure two different job allocations with three executions per allocation. We report the median of the longest-running node among all runs with the corresponding 95\% CI. The reported results do not include the read-in of the graph and the relative distribution phase but only the time taken for the LCC computation.

\begin{table}[h]
    \vspace{-0.5em}
    \renewcommand{\arraystretch}{1.3}
    \centering
    \caption{\review{Graphs used in this paper.}{R4.5}{red}}
    \label{tab:snap_graphs}
    \begin{tabular}{l c c c}
    \hline
    \textbf{Name (type)} & $\mathbf{|V|}$ & $\mathbf{|E|}$ & \textbf{CSR Size}\\
     \hline
    SNAP-Orkut (U) &3 M&117.2 M& 905.8 MiB\\
    SNAP-LiveJournal (U) &4 M&34.7 M& 273.8 MiB\\
    SNAP-LiveJournal1 (D) &4.8 M&69 M& 273.7 MiB\\
    SNAP-Skitter (U) &1.7 M&11.1 M&89.5 MiB\\
    uk-2005 (D) &39.5 M&936.4 M& 3.6 GiB\\
    wiki-en (D)&13,6 M&437.2 M& 1.7 GiB\\
    R-MAT S21 EF16 (U)&2.1 M&33.6 M& 251.1 MiB\\
    R-MAT S23 EF16 (U)&8.4 M&134.2 M& 1021 MiB\\
    R-MAT S30 EF16 (U)&1073.7 M&17179.9 M& 130 GiB\\
    \hline
\end{tabular}
\end{table}

\subsection{Comparison baseline}

\begin{table}[b]
    \vspace{-1em}
    \renewcommand{\arraystretch}{1.3}
    \centering
    \caption{Performance comparison of the different intersection methods on different graphs using 16 threads. We report the number of edges processed per microsecond.}
    \label{tbl:int_comp}
    \begin{tabular}{l c c c}
        \hline
        \textbf{Name} & \textbf{Hybrid} & \textbf{SSI} & \textbf{Binary search}\\
        \hline
        R-MAT S20 EF8 & 0.540 &	0.508 &	0.449\\
        R-MAT S20 EF16 & 0.425 &	0.403 &	0.340\\
        R-MAT S20 EF32 & 0.325 &	0.311 &	0.250\\
        LiveJournal & 1.084 &	1.018 &	0.984\\
        Orkut & 0.596 & 0.552 & 0.503\\
        \hline
    \end{tabular}
\end{table}

\begin{figure}[b]
    \vspace{-0.5em}
    \centering
    \includegraphics[width=\columnwidth]{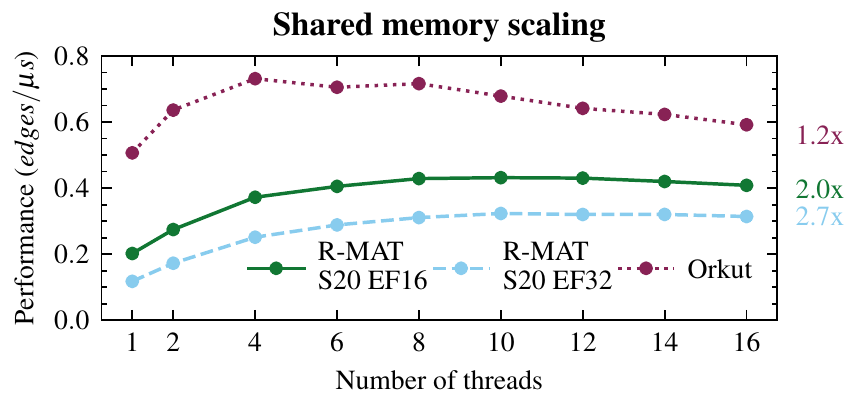}
    \vspace{-2em}
    \caption{Strong scaling on shared memory with hybrid method. We report the number of edges processed per microsecond.}
    \label{fig:int_scale}
\end{figure}

\begin{figure*}[t]
\centering
\includegraphics{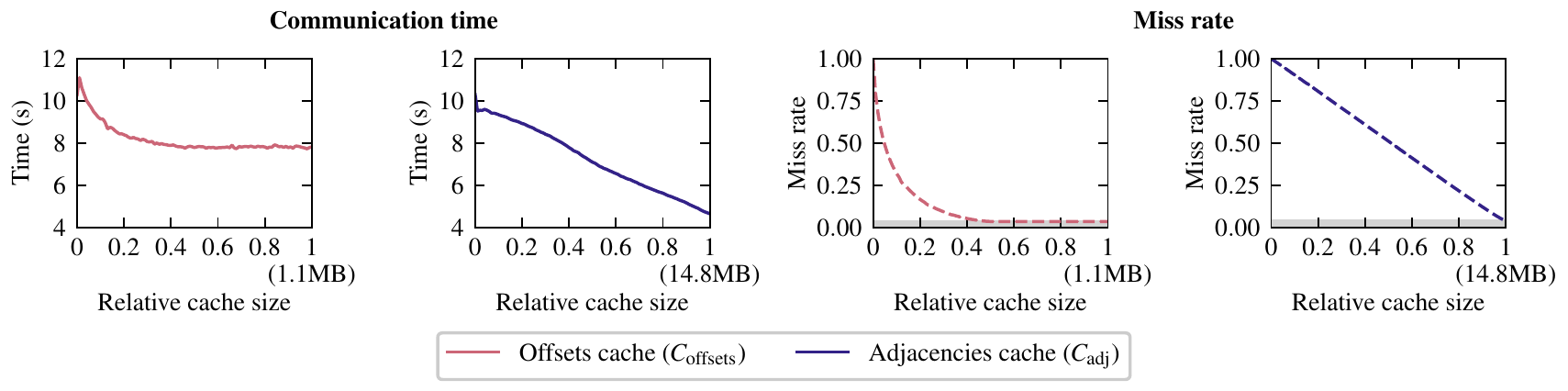}
\caption{Cache behaviour as a function of the cache size. We enable caching only on the respective window, and issue non-cached reads on the other. We used an R-MAT graph with $2^{20}$ vertices and $2^{24}$ edges distributed among 2 computing nodes and measured 100 configurations for both caches. The grey area shows compulsory misses.}
\vspace{-1em}
\label{fig:mem_scale}
\end{figure*}

\review{We compare our solution to TriC~\cite{Tric}, a state-of-the-art, distributed-memory framework for global triangle counting. TriC achieves TC in a per-vertex fashion, implicitly computing LCC scores. The main difference to our solution lies in the query-response approach used by TriC to check for necessary remote edges that leads to synchronization between processes.}{G2}{Green}

\review{TriC's memory demand significantly increases for scale-free graphs, often leading to out-of-memory errors. In those cases, we employed a new version of TriC (\emph{TriC Buffered}) that allocates fixed-size buffers towards remote processes.}

\review{We build TriC using the same Intel ICC 19.1 compiler and map tasks to CPU cores, as it is an MPI-only implementation. For every execution, we specify the \texttt{-b} flag to achieve a more balanced partitioning. We set the buffer size to the largest possible value not bigger than 16 MiB. This cap was necessary due to a network protocol change for messages bigger than 16 MiB that led to higher communication overheads.}

\subsection{Shared memory experiments}

Our measurements comparing the different methods for computing the intersection are summarized in Table~\ref{tbl:int_comp}. The trade-offs between binary search and sorted set intersection were discussed in Hu et al.~\cite{hu2018tricore}. They locate the main weakness of binary search on CPUs in the random accesses in the lookup tree, which leads to a high number of cache misses. On the contrary, SSI traverses the arrays linearly making possible close to zero cache misses. However, in a graph with skewed degree distribution, most edges connect vertices with degrees that are multiple orders of magnitude different from each other. In that case, binary search is essential, as its running complexity is logarithmic in the length of the longer array. Our results show the necessity of both methods on CPU, as the hybrid version always performed better than using SSI or binary search exclusively.

A strong scaling experiment was carried out to measure the gains of computing the intersection in parallel (Figure~\ref{fig:int_scale}). \review{We distribute the local computation on edge level that leads to leaving and re-entering the parallel region for every edge. In effect, the large number of OpenMP library calls becomes a performance bottleneck, which we could also justify by profiling our implementation.}{R1.1}{Blue} \review{We could see a minor improvements of 2\%-4\% with using \texttt{OMP\_WAIT\_POLICY=active}.}{R1.2}{Blue} The best results are achieved for the R-MAT S20 EF32 graph with a speedup of 2.7$\times$ from 1 to 16 threads.

\subsection{Distributed memory experiments}
\subsubsection{Caching performance}

The time of a remote read of size $s$ bytes can be modeled by $t(s) = \alpha + s \cdot \beta$, where $\alpha$ is the setup overhead and $\beta$ the time to read one byte. This means that for the analysis of the cache, both the number of \emph{get}s saved by caching (hit rate) as well as the size of such \emph{get}s have to be taken into consideration.

In Figure~\ref{fig:mem_scale} we demonstrate how the difference between the entries stored in $C_{\textrm{offsets}}$ and $C_{\textrm{adj}}$ (see Section~\ref{sec:datareuse}) and the above remark on the duration of a remote memory read influence communication time for LCC computation. The power law-like relationship between the miss rate and $C_{\textrm{adj}}$'s size is the straight consequence of our algorithm and the graph's degree distribution. In this case, we also notice how already a small memory overhead (relative to $|V|$) allows us to save up to 30\% of the time spent on communication.  In contrast, we observe a linear relationship between the miss rate in $C_{\textrm{offsets}}$ and its size due to the connection between an entry's size and the frequency of remote reads targeting the entry. However, a large portion of transferred data comes from remote reads targeting $w_{\textrm{adj}}$, which is reflected in the reduction in communication time achieved by $C_{\textrm{adj}}$. In this experiment, we reduce communication time by 51.6\% with caching only $w_{\textrm{adj}}$. We expect that for larger datasets, this difference will grow further. Summarizing, with a small memory overhead, one can save an initial amount of communication that targets $w_{\textrm{offsets}}$. However, any further decrease will have a memory overhead linear in terms of the graph's size.

\begin{figure}[t]
    \centering
    \includegraphics[width=\columnwidth]{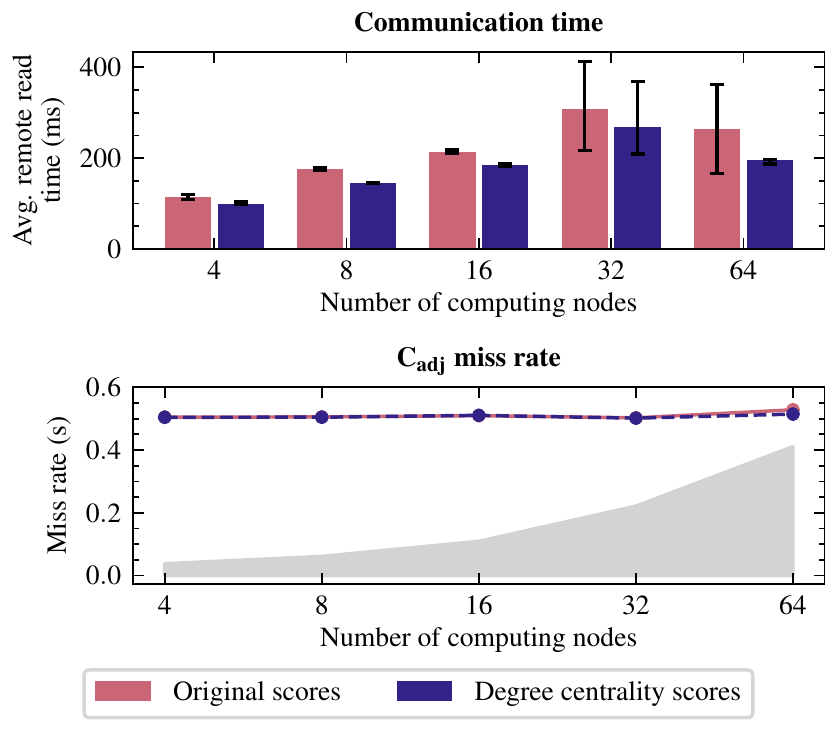}
    \caption{Comparison of original and user-defined scores. We show the average time taken for reading a remote vertex. The grey area shows compulsory misses.}
    \vspace{-1em}
    \label{fig:score_comparison}
\end{figure}

\input{Inlines/overall_performance}

In the following, we assess the effect of the application-specific score described in Section~\ref{sec:datareuse}. In this experiment, we fix the memory allocated for $C_{\textrm{adj}}$ to 25\% of the size of the graph's non-local partition at every node to trigger the eviction procedure. The results for an R-MAT graph with $2^{20}$ vertices and $2^{24}$ edges are shown in Figure~\ref{fig:score_comparison}. Degree centrality scores improve caching performance between 14.4\% and 35.6\% with respect to the original scores for this dataset.


\subsubsection{Overall performance}

Finally, in Figures~\ref{fig:full_exp} and~\ref{fig:big_exp} we report the performance of our LCC implementation both with and without caching. Next to the measurements, we denote the speedup achieved with respect to the smallest configuration. \review{For experiments using caching, a total of \emph{16 GiB} memory is reserved, allocating $0.8 \cdot |V|$ bytes for $C_{\textrm{offsets}}$ and the rest for $C_{\textrm{adj}}$.}{R3.10}{orange} \review{Note, that with this configuration $C_{\textrm{offsets}}$ can store  $0.4 \cdot |V|$ many vertices, as the position of a remote adjacency list is given as a pair of \emph{(start, end)} positions.}{R3.11}{orange} Furthermore, we also show measurements using TriC to better asses our implementation. \review{In case of missing data points, the corresponding experiment exceeded a wall-time of 9 hours.}{G2}{Green}

As the graph is distributed among more computing nodes, the number of edges that cause communication increases. For example, for the R-MAT S21 EF16 graph, the average fraction of remote and local reads grows from 66\% to 98\%. For the same graph, with 4 computing nodes, communication already takes up to 78.9\% of the total running time, which rises to 97.7 \% for 64 nodes. In general, we could observe that communication quickly dominates the total running time, implying that the limitations of the shared memory computation have minor effects on overall performance. \review{Moreover, it follows that overlapping communication and computation, even for more than a single edge, cannot significantly improve the runtime.}{R3.5}{orange} As communication dominates, the speedup achievable is determined by the number of edges causing communications. At small scale, the worst scaling for a real-world graph is observed on the Orkut graph with $9.4\times$ speedup from 4 to 64 nodes. We explain this with the following two reasons. Firstly, for this graph, the running time is mostly determined by computation when 4 nodes are used, but quickly becomes communication-bound with larger configurations. Secondly, in this case, 1D partitioning causes load imbalance, leading to an up 25\% difference in the running time of the different processes. We achieve our best results for the LiveJournal1 graph, with 14$\times$ speedup from 4 to 64 computing nodes. Our large-scale experiments are similarly limited by load imbalance, which bounds the achieved scaling.

\footnotetext{We report only one measurement per data point for the RMAT S30 EF16 graph due to cost reasons.}

\review{The efficiency of caching is influenced by multiple factors. Firstly, in case the cache size is significantly smaller than the size of the graph, capacity misses are unavoidable. Secondly, the graph structure inherently determines the possible data reuse, and graphs with flatter degree distribution will lead to a large number of compulsory misses.}{R4.6}{Red} Finally, by distributing the graph among an increasing number of computing nodes, data reuse reduces due to the increasing number of edges that cross partitions. This will similarly result in an increased number of compulsory misses. Compulsory misses decrease the overall hit rate and incur an overhead caused by the caching process in CLaMPI. For example, for the LiveJournal graph, 15.5\% of the remote reads are compulsory misses for 4 nodes (a compulsory miss is always a compulsory miss in both caches) that grows up to 64.9\% with 64 nodes. 

Regarding the overall performance, we can distinguish between 3 scenarios: (1) computation dominates and caching has no significant effect; (2) high number of compulsory misses limit caching performance; (3) caching is beneficial and reduces the communication time. Scenario (1) can be seen for the Orkut graph where the effect of caching increases from 4 to 8 computing nodes. Scenario (2) is especially notable for the LiveJournal and LiveJournal1 graphs. In these cases, CLaMPI's overhead leads to worse performance than the non-cached version. However, we see significant reduction in running times between these two extremes. At small scale, we achieve up to 67\% and 47\% better running times for the R-MAT S21 EF16 and LiveJournal graphs, respectively. \review{At large scale, the cached version resulted in 73\% better performance compared to the non-cached implementation for the R-MAT S30 EF16 graph. We remark, that this result is achieved with a cache size of only 12\% of the graph's CSR representation.}{G1}{Green}

\review{We achieve significantly better execution times both at large-scale (up to 3.6x speedup) and at small-scale (up to 100x) compared to TriC. The advantages of our asynchronous implementation over TriC is especially notable with the synthetic datasets that posses a close to perfect scale free degree distribution.}{G2}{Green} Despite the imbalance between processes coming from the 1D distribution, we conclude that these results justify the necessity of an asynchronous algorithm for distributed-memory LCC computation. 

\section{Related Work}
In the following, we summarize the main techniques used for shared and distributed-memory TC analysis. For a thorough comparison of the different triangle counting algorithms, we refer the reader to the paper from Schank and Wagner~\cite{tclisting}, and to work by Shun et al.~\cite{shun2015multicore}.

\subsection{Frontier intersection}
SSI has been introduced by Green et al.~\cite{intersect-path}, and binary search appeared first for triangle counting in Hu et al.~\cite{hu2018tricore}. Pandey et al.~\cite{pandey2019h} utilize hashing for computing intersections but, instead of hashing every element, they use a selected number of bins where multiple elements are stored. This solution was further improved in their recent work~\cite{TRUST}.

\subsection{Algebraic computation}

For a graph $G$ one can compute the matrix $\mathbf{C} = \mathbf{A}\mathbf{A}\circ\mathbf{A}$, whose entry $c_{ij}$ stores the number of triangles that contain $e_{ij}$. For undirected graphs, this can be simplified to $\mathbf{C} = \mathbf{L}\mathbf{U}\circ\mathbf{A}$, where $\mathbf{L}$ and $\mathbf{U}$ are the lower and upper triangular matrices. Triangle counting implementations based on this algebraic computation method take advantage of the sparsity of $G$ and use highly optimized libraries for sparse matrix multiplication. An improved parallel implementation can be found in the paper from \review{Azad et al.~\cite{azad2015parallel}}{R1.5}{Blue}, and a distributed algebraic-based TC algorithm has been proposed by Hutchinson~\cite{hutchison2017distributed} using the Apache Accumulo distributed database. Aznaveh et al.~\cite{graphblas} implemented shared memory parallel TC and LCC computation based on the SuiteSparse\:GraphBLAS implementation of the GraphBLAS standard.

\subsection{Distribution techniques}
For any TC or LCC algorithm that utilizes parallelism on some level, work distribution is of primary importance. Kolda et al.~\cite{kolda2014counting} use the Mapreduce technique for triangle counting. Lumsdaine et al.~\cite{CyclicDistribution} introduced a cyclic distribution for 1D to achieve balanced partitions. Two-dimensional partitioning assigns edges to processes in a grid-based manner. \reversemarginpar \review{Tom and Karypis~{\cite{tom20192d}}}{R1.4}{Blue}\normalmarginpar~developed a triangle counting algorithm for undirected graphs following a parallel matrix multiplication scheme based on 2D. Acer et al.~\cite{8916302} utilize 2D partitioning among the computing nodes and achieves shared memory parallelism based on 1D. Hoang et al.~\cite{hoang2019disttc} compute shadow edges and corresponding vertices that are necessary for local triangle computation, thus avoiding any communication during the computation phase. We emphasize that all the aforementioned work requires synchronization mechanisms, and therefore, their scalability is limited.

\section{Conclusion}
We introduce a fully asynchronous distributed-memory algorithm for both triangle counting and LCC. Synchronization overheads are removed by using RMA one-sided operations to retrieve remote parts of the graph that are needed to progress the algorithm (i.e., parts of the adjacency lists that have been partitioned and assigned to remote peers). Additionally, we show how irregular graph algorithms such as LCC and TC expose data reuse, which we exploit by using a transparent caching solution for RMA, i.e., CLaMPI. To improve cache efficiency, we extend CLaMPI to take into account application-specific scores when deciding which entries to evict in case of conflict or capacity misses. For example, by using degree centrality as the score for LCC, we are able to reduce the total running time by up to \review{73\%}{G1}{Green}. Overall, we show that removing synchronization costs and achieving vertex delegation by a caching mechanism leads to clear performance improvements over the current state-of-the-art.
Finally, we plan to extend this work in many directions by
i) designing new asynchronous algorithms for TC/LLC based on distribution schema that have lower communication costs than 1D distribution~\cite{10.1007/978-3-642-23397-5_10}; ii) investigating other graph problems that may benefit from the proposed approach~\cite{9432723, 10.1145/3182656, solomonik2017scaling, besta2020high} and, in general, those that can be expressed in a push-pull dichotomy~\cite{10.1145/3078597.3078616}; iii) studying other application-specific scores for cached entries to improve caching efficiency. 

\section*{Acknowledgment}

\footnotesize This work has been partially funded by the UNIBZ-RTD-CALL2018-IN2087, INdAM–GNCS 2020-NoRMA, MIU-FISR-2020-FISR2020IP\_00802 projects, and the European Project RED-SEA (Grant No.~955776). 
This project has received funding from the European Research Council (ERC) under the European Union’s Horizon 2020 research and innovation program (Grant agreement No.~101002047).
We thank the Swiss National Computing Center (CSCS) for providing computing resources and excellent technical support.

\begin{figure}[h]
    \centering
    \includegraphics[height=3em]{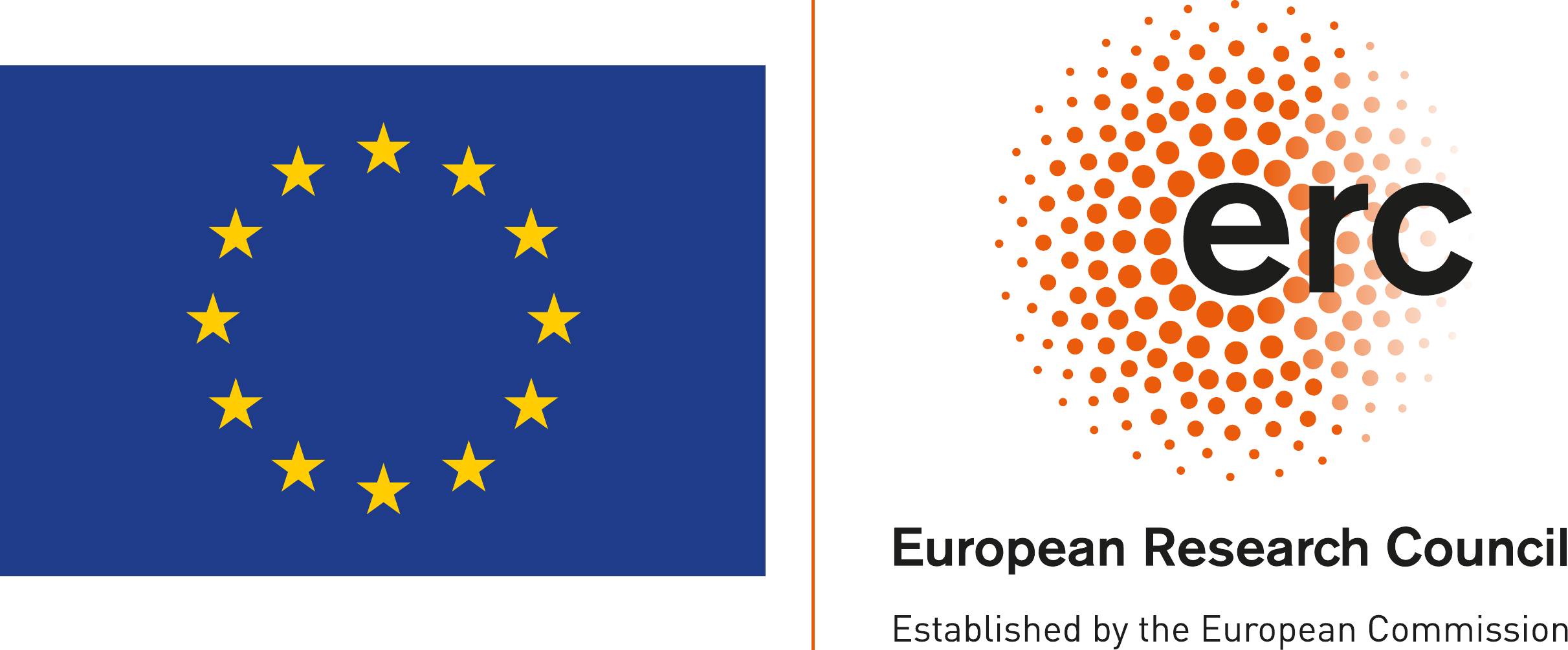}
\end{figure}

\bibliographystyle{IEEEtran}
\bibliography{refs}

\end{document}

%% file: Inlines/symbols.tex
\begin{table}[t]
\centering
\footnotesize
\caption{Symbols used in the paper.}
\begin{tabular}{@{}ll@{}}
\toprule
                    $G=(V,E)$&A graph $G$ where $V$, $E$ are sets of vertices and edges.\\
                    $n$&number of vertices.\\
                    $deg^+(v_i)$&out-degree of $v_i$.\\
                    $deg^-(v_i)$&in-degree of $v_i$.\\
                    $adj(v_i)$& adjacency of vertex $v_i$.\\ 
                    $\mathbf{A}$& adjacency matrix of $G$.\\
                    $\triangle_{ijk}$& a triangle including the edges $e_{ij}, e_{jk}, e_{ik} \in E$\\
                    $p$& number of computing nodes\\
\bottomrule
\end{tabular}
\label{tab:symbols}
\end{table}

%% file: Inlines/bin_search.tex
\begin{algorithm}[!t]
\footnotesize
\caption{Binary Search for $A \cap B$}
\label{algo:bin_search}
\begin{algorithmic}[1]
\Function{BinarySearch}{$A,B$} 

    \State $counter, bottom \leftarrow 0$
    \State $top \leftarrow |B| - 1$
    \ForAll{$x \in A$}
    \While{$bottom < top - 1$}
    \State $mid = \lfloor(top - bottom) / 2\rfloor$
    \If{$x < B[mid]$} 
    \State $top \leftarrow mid$
    \ElsIf{$x > B[mid]$}
    \State $bottom \leftarrow mid$
    \Else
    \State $counter \leftarrow counter + 1$
    \State \textbf{break}
    \EndIf
    \EndWhile
    \EndFor
    
    \Return{$counter$}
    
\EndFunction

\end{algorithmic}
\end{algorithm}

%% file: Inlines/ssi.tex
\begin{algorithm}[!t]
\footnotesize
\caption{Sorted Set Intersection for $A \cap B$}
\label{algo:ssi}
\begin{algorithmic}[1]
\Function{SSI}{$A,B$} 
    \State $counter \leftarrow 0$
    \State $i,j \leftarrow 0$
    \While{$i < |A|$ \textbf{and} $j < |B|$}
    \If{$A[i] == B[j]$} 
    \State $counter \leftarrow counter + 1$
    \State $i\leftarrow i + 1$ 
    \State $j\leftarrow j + 1$
    \ElsIf{$A[i] < B[j]$}
    \State $i\leftarrow i + 1$
    \Else
    \State $j\leftarrow j + 1$
    \EndIf
    \EndWhile
    
    \Return{$counter$}
    
\EndFunction
\end{algorithmic}
\end{algorithm}

%% file: Inlines/lcc_comp.tex
\begin{algorithm}[t]
\footnotesize
\caption{\review{Distributed LCC Computation}{R4.3}{red}}
\label{algo:dist_lcc}
\begin{algorithmic}[1]
\Procedure{Distributed LCC}{}
\State Exchange vertices based on 1D partitioning
\State Build CSR representation
\ForAll{locally owned vertex $v_i$}
\State $t \leftarrow 0$ \Comment{Stores the number of triangles.}
\ForAll{$e_{ij}$ such that $v_j \in adj(v_i)$}
\If{$v_j$ is remote}
\State RemoteRead($adj(v_j)$) \Comment{See Sec.~\ref{sec:datareuse}}
\EndIf
\State $t \mathrel{+}=$ Intersect($adj(v_i), adj(v_j)$) \Comment{See Sec.~\ref{sec:dynamic}} 

\EndFor
\State Compute LCC score of $v_i$ with $t$ triangles.
\EndFor
\EndProcedure
\end{algorithmic}
\end{algorithm}

%% file: Inlines/overall_performance.tex
\begin{figure*}[t]
\normalsize
\centering
\includegraphics{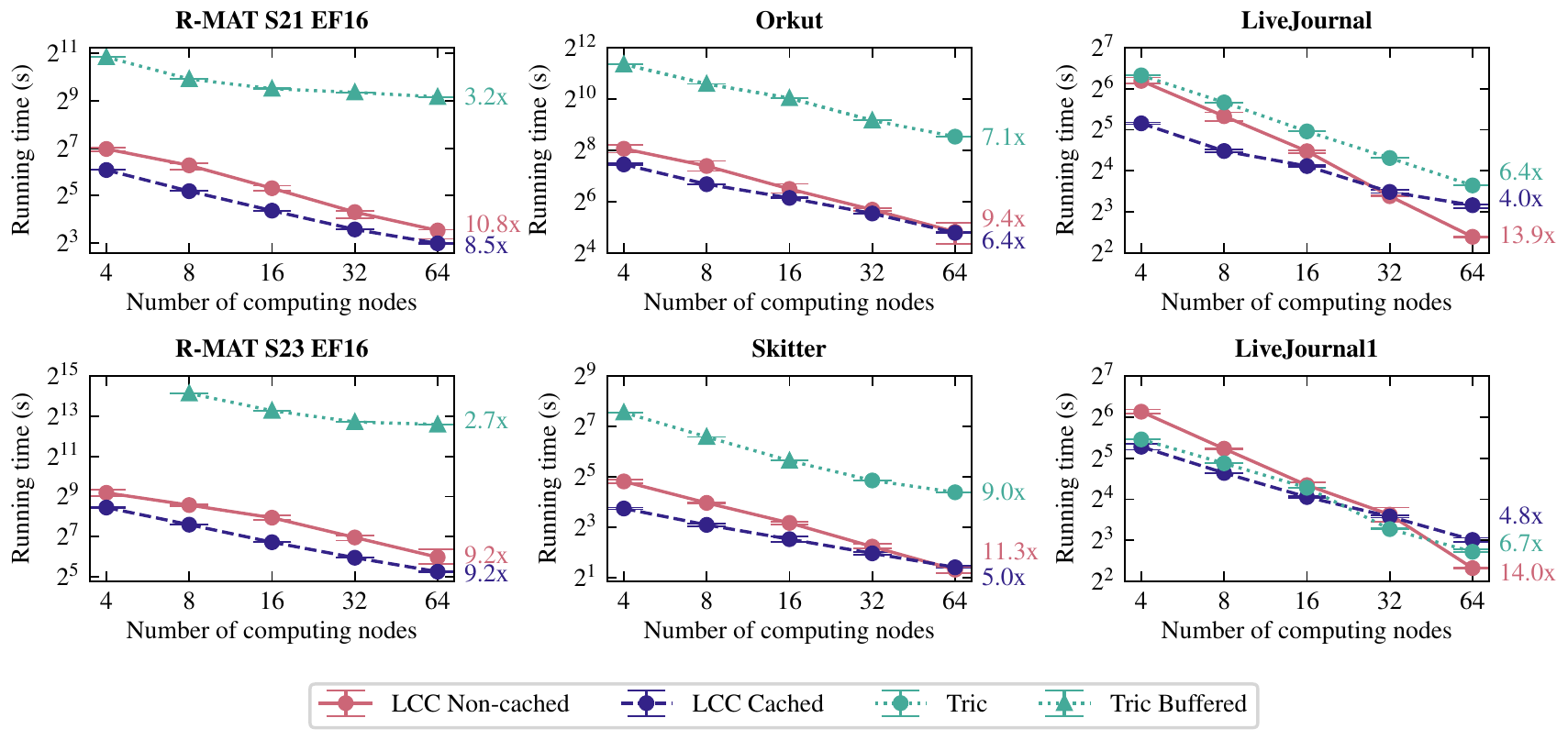}
\caption{\review{Strong scaling experiments on small scale with \emph{16 GiB} memory overhead (log-log scale).}{G2}{Green}}
\label{fig:full_exp}
\vspace{1em}
\includegraphics{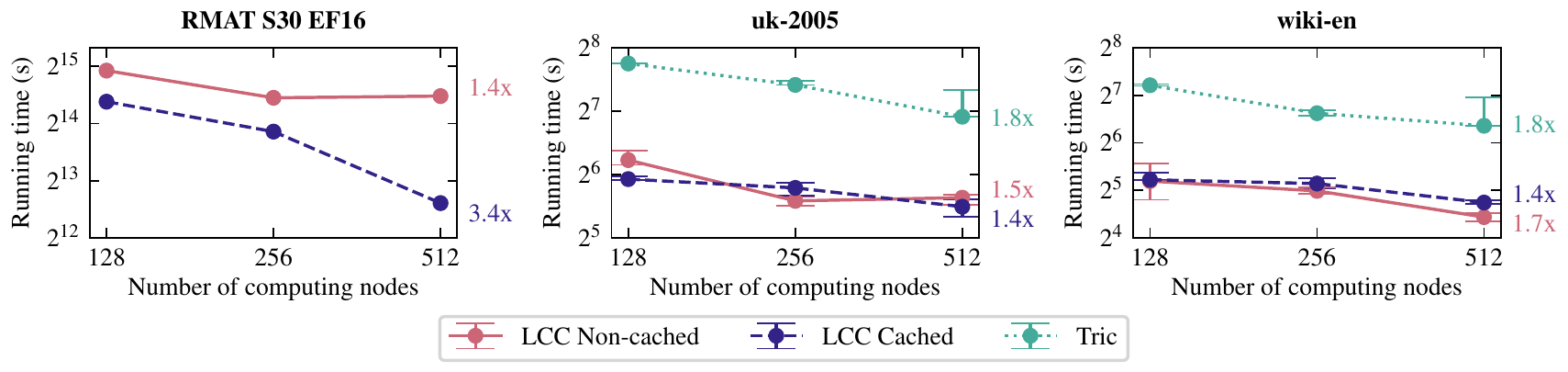}
\caption[Caption]{Strong scaling experiments on large scale with \emph{16 GiB} memory overhead (log-log scale).\footnotemark}
\label{fig:big_exp}
\vspace{-1em}
\end{figure*}